# On the accuracy of the state space restriction approximation for spin dynamics simulations

Alexander Karabanov[1,*], Ilya Kuprov[2,*], G.T.P. Charnock[2], Anniek van der Drift[1]
Luke J. Edwards[2], Walter Köckenberger[1]

[1]*Sir Peter Mansfield Magnetic Resonance Centre, School of Physics & Astronomy, University of Nottingham, University Park, Nottingham, NG7 2RD, UK.*

[2]*Oxford e-Research Centre, University of Oxford,
7 Keble Road, Oxford OX1 3QG, UK.*

[*]Corresponding Authors:

alexander.karabanov@nottingham.ac.uk (AK)

ilya.kuprov@oerc.ox.ac.uk (IK)

**Abstract**

We present an algebraic foundation for the state space restriction approximation in spin dynamics simulations and derive applicability criteria as well as minimal basis set requirements for practically encountered simulation tasks. The results are illustrated with NMR, ESR, DNP and Spin Chemistry simulations. It is demonstrated that state space restriction yields accurate results in systems where the time scale of spin relaxation processes approximately matches the time scale of the experiment. Rigorous error bounds and basis set requirements are derived.

# 1. Introduction

Time-domain simulation methods in magnetic resonance spectroscopy have recently made considerable progress – polynomially scaling algorithms have been published[1-4] and shown to work[5-8] for the simulation of several classes of spin dynamics processes and magnetic resonance experiments. For specific types of NMR simulations, linear scaling algorithms have recently emerged[4,9]. This marks a significant improvement on the exponential scaling situation of just a few years ago. Several software packages taking advantage of these polynomially scaling algorithms have recently been released[4,7,10,11].

All methods proposed so far use the concept of *state space restriction*, which is based on the assumption that the complete basis set is not necessary, because large spin systems do often only occupy a fraction of their state spaces during time evolution[1,2,8]. This argument is not constrained to (or derived from) the short-time approximation[12], which has its roots in perturbation theory – the subspaces that a typical spin system trajectory would not populate stem from symmetries and conservation laws[8] as well as (*vide infra*) relaxation behaviour.

An important topic that we seek to address in this communication is the general algebraic analysis of the state space restriction approximation for spin dynamics simulations and the derivation of rigorous applicability and validity conditions. This paper derives such conditions and provides a discussion of their practical consequences.

# 2. Applicability ranges for state space restriction

We will start with the Liouville space representation of the master equation for the dynamics of the density matrix $\hat{\rho}$:

$$\frac{\partial}{\partial t}\hat{\rho} = -i\hat{\hat{H}}\hat{\rho} + \hat{\hat{R}}(\hat{\rho} - \hat{\rho}_{eq}), \qquad \hat{\hat{H}} = ad_{\hat{H}} = [\hat{H},\ ], \qquad \hat{\hat{H}} = \hat{\hat{H}}_1 + \hat{\hat{H}}_2 \qquad (1)$$

in which $\hat{H}$ is the Hamiltonian, $\hat{\hat{R}}$ is the relaxation superoperator and $\hat{\rho}_{eq}$ is the thermal equilibrium density matrix. To facilitate the subsequent treatment, the Hamiltonian will be partitioned into single-spin terms (Zeeman, quadrupolar, zero-field splitting, radiofrequency, microwave), denoted collectively $\hat{H}_1$, and two-spin terms (all spin-spin couplings) denoted $\hat{H}_2$.

The state space of the spin system may be represented as a direct sum of subspaces spanned by operators corresponding to correlations between specific numbers of spins:

$$\mathfrak{L} = \mathfrak{L}_0 \oplus \mathfrak{L}_1 \oplus \mathfrak{L}_2 \oplus \ldots \oplus \mathfrak{L}_N \qquad (2)$$

where $N$ is the number of spins and $\mathfrak{L}_k$ is the subspace of $k$-spin correlations, spanned (for example) by direct products of $k$ irreducible spherical tensor operators each acting on its own spin. The $\mathfrak{L}_0$ subspace is spanned by the unit operator. Each subspace $\mathfrak{L}_k$ is closed under $\hat{\hat{H}}_1$

$$\hat{\hat{H}}_1 \mathfrak{L}_k \subset \mathfrak{L}_k \tag{3}$$

because the correlation order of a given state cannot be changed by taking a commutator with a single-spin operator. $\mathfrak{L}_k$ does, however, leak into adjacent subspaces under $\hat{\hat{H}}_2$:

$$\hat{\hat{H}}_2 \mathfrak{L}_k \subset \mathfrak{L}_{k-1} \oplus \mathfrak{L}_k \oplus \mathfrak{L}_{k+1} \tag{4}$$

because a commutator of a two-spin operator with a $k$-spin operator can increase or reduce the correlation order by one spin, as well as leave it unchanged, for example:

$$[\hat{L}_Z \hat{S}_Z, \hat{S}_+ \hat{I}_-] = \hat{L}_Z \hat{S}_+ \hat{I}_-, \quad [\hat{L}_Z \hat{S}_Z, \hat{L}_Z \hat{S}_+ \hat{I}_-] = \begin{cases} (1/4)\hat{S}_+ \hat{I}_- & \text{for spin } 1/2 \\ \hat{L}_Z^2 \hat{S}_+ \hat{I}_- & \text{for spin} > 1/2 \end{cases} \tag{5}$$

A similar split with similar arguments may be applied to the relaxation superoperator:

$$\hat{\hat{R}}_1 \mathfrak{L}_k \subset \mathfrak{L}_k \qquad \hat{\hat{R}}_2 \mathfrak{L}_k \subset \mathfrak{L}_{k-2} \oplus \mathfrak{L}_{k-1} \oplus \mathfrak{L}_k \oplus \mathfrak{L}_{k+1} \oplus \mathfrak{L}_{k+2} \tag{6}$$

where the negative-definite $\hat{\hat{R}}_1$ term governs longitudinal and transverse relaxation as well as cross-relaxation within the same subspace, whereas the indefinite $\hat{\hat{R}}_2$ term is responsible for cross-correlations[13,14]. If Bloch-Redfield-Wangsness theory is used to describe relaxation[15-17], $\hat{\hat{R}}_2$ can mix $\mathfrak{L}_k$ with $\mathfrak{L}_{k\pm 2}$ due to the presence of a double commutator in the $[\hat{H}(t),[\hat{H}(t'),\hat{\rho}(t)-\hat{\rho}_{eq}]]$ term at the core of that theory. The cross-correlation processes are slower than the self-relaxation processes in $\hat{\hat{R}}_1$, and for the purpose of obtaining bounds on the overall density matrix norm they are likely to be non-essential. In the treatment below we shall therefore only keep the $\hat{\hat{R}}_1$ term.

In common experimental practice, the dynamics starts in $\mathfrak{L}_1$ (NMR, ESR, DNP, *etc.*) or $\mathfrak{L}_2$ (PHIP, CIDNP, CIDEP, *etc.*) and is detected either with magnetization operators from $\mathfrak{L}_1$ or with singlet and triplet operators from $\mathfrak{L}_2$. With the exception of very low temperature experiments, the thermal equilibrium state is also in $\mathfrak{L}_1$. We are therefore interested in finding accurate representations for spin dynamics around $\mathfrak{L}_1$ and in determining the extent to which higher spin orders are required to achieve a user-specified accuracy target.

With the state space partitioned according to Equation (2) the density matrix $\hat{\rho}$ may also be split up into contributions from the spin correlations involving specific numbers of spins:

$$\hat{\rho} = \hat{\rho}_0 + \hat{\rho}_1 + \hat{\rho}_2 + \ldots + \hat{\rho}_N \qquad \hat{\rho}_k \in \mathfrak{L}_k \qquad \hat{\rho}_{eq} \in \mathfrak{L}_1 \tag{7}$$

which are connected by the following system of equations derived from Equation (1):

$$\begin{cases} \dfrac{\partial}{\partial t}\hat{\rho}_1 = -\mathrm{i}\hat{\hat{H}}_1\hat{\rho}_1 - \mathrm{i}\hat{\hat{\pi}}_1\hat{\hat{H}}_2(\hat{\rho}_1 + \hat{\rho}_2) + \hat{\hat{R}}_1(\hat{\rho}_1 - \hat{\rho}_{\mathrm{eq}}) \\ \dots \\ \dfrac{\partial}{\partial t}\hat{\rho}_k = -\mathrm{i}\hat{\hat{H}}_1\hat{\rho}_k - \mathrm{i}\hat{\hat{\pi}}_k\hat{\hat{H}}_2(\hat{\rho}_{k-1} + \hat{\rho}_k + \hat{\rho}_{k+1}) + \hat{\hat{R}}_1\hat{\rho}_k \\ \dots \\ \dfrac{\partial}{\partial t}\hat{\rho}_N = -\mathrm{i}\hat{\hat{H}}_1\hat{\rho}_N - \mathrm{i}\hat{\hat{\pi}}_N\hat{\hat{H}}_2(\hat{\rho}_{N-1} + \hat{\rho}_N) + \hat{\hat{R}}_1\hat{\rho}_N \end{cases} \qquad (8)$$

where $\hat{\hat{\pi}}_k$ is the superoperator projecting density matrices into $\mathfrak{L}_k$. The first equation is special in that it includes the thermal equilibrium term $\hat{\rho}_{\mathrm{eq}} \in \mathfrak{L}_1$, which is replenishing $\mathfrak{L}_1$ *via* the relaxation superoperator. After we note that

$$\hat{\hat{\pi}}_k \hat{\hat{H}}_2 (\hat{\rho}_{k-1} + \hat{\rho}_k + \hat{\rho}_{k+1}) = \hat{\hat{\pi}}_k \hat{\hat{H}}_2 (\hat{\rho}_0 + \hat{\rho}_1 + \hat{\rho}_2 + \dots + \hat{\rho}_N) = \hat{\hat{\pi}}_k \hat{\hat{H}}_2 \hat{\rho}$$

$$\hat{\hat{\pi}}_1 + \hat{\hat{\pi}}_2 + \dots + \hat{\hat{\pi}}_N = \hat{\hat{E}} \qquad (9)$$

where $\hat{\hat{E}}$ is the identity superoperator, it becomes easy to demonstrate that the vertical sum of all equations for the components of $\hat{\rho}$ in Equations (8) is equal to the master equation and therefore Equations (8) are just a re-formulation of Equation (1) with the subspace partitioning clearly exposed for analysis.

Our objective in analyzing these recurrence relations is to determine the extent to which an accurate representation of dynamics in $\mathfrak{L}_1$ would require the inclusion of subspaces containing higher spin operators. The extent to which the system occupies a particular subspace $\mathfrak{L}_k$ is given by the norm of the corresponding part of the density matrix $\hat{\rho}_k$. The differential equations for these norms may be obtained directly from Equations (8):

$$\frac{\partial}{\partial t}|\hat{\rho}_k|^2 = \frac{\partial}{\partial t}\langle\hat{\rho}_k|\hat{\rho}_k\rangle = \frac{\partial}{\partial t}\mathrm{Tr}(\hat{\rho}_k^\dagger \hat{\rho}_k) = \mathrm{Tr}\left[\left(\frac{\partial\hat{\rho}_k}{\partial t}\right)^\dagger \hat{\rho}_k + \hat{\rho}_k^\dagger \left(\frac{\partial\hat{\rho}_k}{\partial t}\right)\right], \qquad (10)$$

After substituting the derivatives from Equations (8) and some simplifications, taking into account the fact that $\mathrm{Tr}(\hat{\rho}_k^\dagger \hat{\hat{\pi}}_k \hat{\rho}) = \mathrm{Tr}(\hat{\rho}_k^\dagger \hat{\rho}_k) = |\hat{\rho}_k|^2$, we get the following system of equations for the norms (Dirac notation for matrix scalar products is used for clarity):

$$\begin{cases} \dfrac{\partial}{\partial t}|\hat{\rho}_1|^2 = -2\operatorname{Im}\left(\langle\hat{\rho}_2|\hat{\hat{H}}_2|\hat{\rho}_1\rangle\right) + 2\langle\hat{\rho}_1|\hat{\hat{R}}_1|\hat{\rho}_1\rangle - 2\operatorname{Re}\left(\langle\hat{\rho}_1|\hat{\hat{R}}_1|\hat{\rho}_{\text{eq}}\rangle\right) \\ \ldots \\ \dfrac{\partial}{\partial t}|\hat{\rho}_k|^2 = -2\operatorname{Im}\left(\langle\hat{\rho}_{k+1}|\hat{\hat{H}}_2|\hat{\rho}_k\rangle\right) + 2\operatorname{Im}\left(\langle\hat{\rho}_k|\hat{\hat{H}}_2|\hat{\rho}_{k-1}\rangle\right) + 2\langle\hat{\rho}_k|\hat{\hat{R}}_1|\hat{\rho}_k\rangle \\ \ldots \\ \dfrac{\partial}{\partial t}|\hat{\rho}_N|^2 = 2\operatorname{Im}\left(\langle\hat{\rho}_N|\hat{\hat{H}}_2|\hat{\rho}_{N-1}\rangle\right) + 2\langle\hat{\rho}_N|\hat{\hat{R}}_1|\hat{\rho}_N\rangle \end{cases} \quad (11)$$

Because the eigenvalue ranges of both the Hamiltonian and the relaxation superoperator are bounded, the following relations must hold (*eig* refers to the set of eigenvalues):

$$\begin{aligned} \operatorname{Im}\left(\langle\hat{\rho}_{k+1}|\hat{\hat{H}}_2|\hat{\rho}_k\rangle\right) &= h_k|\hat{\rho}_{k+1}||\hat{\rho}_k|, & \min\left|\operatorname{eig}\left(\hat{\hat{H}}_2\right)\right| &\leq |h_k| \leq \max\left|\operatorname{eig}\left(\hat{\hat{H}}_2\right)\right| \\ \langle\hat{\rho}_k|\hat{\hat{R}}_1|\hat{\rho}_k\rangle &= -r_k|\hat{\rho}_k|^2, & \min\left|\operatorname{eig}\left(\hat{\hat{R}}_1\right)\right| &\leq |r_k| \leq \max\left|\operatorname{eig}\left(\hat{\hat{R}}_1\right)\right| \\ \operatorname{Re}\left(\langle\hat{\rho}_1|\hat{\hat{R}}_1|\hat{\rho}_{\text{eq}}\rangle\right) &= -r_0|\hat{\rho}_1||\hat{\rho}_{\text{eq}}|, & \min\left|\operatorname{eig}\left(\hat{\hat{R}}_1\right)\right| &\leq |r_0| \leq \max\left|\operatorname{eig}\left(\hat{\hat{R}}_1\right)\right| \end{aligned} \quad (12)$$

where the negative-definiteness of the relaxation superoperator has been exposed explicitly with a minus sign. Without loss of generality we can scale the problem to $|\hat{\rho}_{\text{eq}}|=1$. With Equations (12) in place, after noting that $\partial|\hat{\rho}_k|^2/\partial t = 2|\hat{\rho}_k|(\partial|\hat{\rho}_k|/\partial t)$ and some simplifications, we get:

$$\begin{cases} \dfrac{\partial}{\partial t}|\hat{\rho}_1| = -h_1|\hat{\rho}_2| - r_1|\hat{\rho}_1| + r_0 \\ \ldots \\ \dfrac{\partial}{\partial t}|\hat{\rho}_k| = -h_k|\hat{\rho}_{k+1}| + h_{k-1}|\hat{\rho}_{k-1}| - r_k|\hat{\rho}_k| \\ \ldots \\ \dfrac{\partial}{\partial t}|\hat{\rho}_N| = h_{N-1}|\hat{\rho}_{N-1}| - r_N|\hat{\rho}_N| \end{cases} \quad (13)$$

For a two-subspace special case it is easy to see that these are not chemical kinetics type equations, but the driven and damped oscillation type equations

$$\begin{cases} \dfrac{\partial}{\partial t}|\hat{\rho}_1| = -h_1|\hat{\rho}_2| - r_1|\hat{\rho}_1| + r_0 \\ \dfrac{\partial}{\partial t}|\hat{\rho}_2| = h_1|\hat{\rho}_1| - r_2|\hat{\rho}_2| \end{cases} \quad (14)$$

which are reasonably expected to arise from the essentially rotational dynamics of the original master equation. Equations (13) expose the hierarchical structure of the spin state space, which is illustrated in Figure 1 – higher spin orders are replenished by the magnetization arriving from below and drained by relaxation. Clearly, in a very large spin system, a kind of equilibrium ought to emerge, wherein the supply would be balanced by the drain and the upward movement of population probability through the spin correlation hierarchy would be halted.

Except for neglecting the cross-correlated relaxation processes, we have not made any approximations in deriving Equations (13). To make progress with estimating the parameters of the above noted equilibrium, we will now make reasonable assumptions about the terms they contain. Because we seek to obtain the *upper bound* for the error introduced by the state space restriction approximation, it would be an appropriate simplification to replace all $h_k$ terms that pump the magnetization up the diagram in Figure 1 by the term having the largest magnitude, keeping the relative signs intact:

$$h_k \quad \rightarrow \quad h = \max_k \{|h_k|\} \leq \left\| \hat{\hat{H}}_2 \right\| \qquad (15)$$

It would similarly be reasonable to replace all $r_k$ terms moving the magnetization in the downward direction with the term having the smallest magnitude. It is, however, known that high-spin orders, do in general relax faster than the low-spin orders[18,19]. It is reasonable to assume that the relaxation rate is approximately proportional to the number of correlated spins:

$$r_k \quad \rightarrow \quad kr = k \min_k \{|r_k|\} \leq \left\| \hat{\hat{R}}_1 \right\| \qquad (16)$$

For liquid-state NMR and ESR systems this assumption is also supported by direct inspection – Figure 2 shows the maximum and the RMS absolute eigenvalues of the projection of $\hat{\hat{R}}$ into a particular subspace $\mathfrak{L}_k$ as a function of the rank of that subspace $k$.

The scaling law assumed in Equation (16) has notable exceptions. In liquid state systems, the singlet states studied recently by Levitt and co-authors[20,21] exhibit very slow relaxation and, if populated, would not be easily drained. However, the same mechanism that makes these states relax slowly (dipolar transitions are permutation symmetry forbidden, Zeeman frequencies match and external couplings are absent or switched off) also makes these states hard to populate – unless specific effort is made to steer the system into a singlet, these states are unlikely to contribute significantly to the density matrix norm. The second exception is the solid state systems studied recently by Krojanski, Lovrić and Suter[22-25]: they found that the relaxation rates scale as the square root of the coherence order (see *e.g.* Figure 6 in Reference [24]). In the treatment below we shall therefore also discuss the case where

$$r_k \quad \rightarrow \quad r\sqrt{k} = \sqrt{k} \min_k \{|r_k|\} \leq \left\| \hat{\hat{R}}_1 \right\| \qquad (17)$$

and provide (skipping the almost identical derivation) the corresponding error bounds.

The error estimates with the simplifications outlined above would be an upper bound for the actual reduced state space approximation error. With Equations (15) and (16) in place, Equations (13) become:

$$\begin{cases} \frac{\partial}{\partial t}|\hat{\rho}_1| = -h|\hat{\rho}_2| - r|\hat{\rho}_1| + r_0 \\ \ldots \\ \frac{\partial}{\partial t}|\hat{\rho}_k| = -h(|\hat{\rho}_{k+1}| - |\hat{\rho}_{k-1}|) - kr|\hat{\rho}_k| \\ \ldots \\ \frac{\partial}{\partial t}|\hat{\rho}_N| = h|\hat{\rho}_{N-1}| - Nr|\hat{\rho}_N| \end{cases} \qquad (18)$$

where $h \geq 0$, $r > 0$, $r_0 > 0$.

In the limit of a very large spin system, the $k$ index may be assumed to be continuous. After taking this limit, we get the following partial differential equation for the transport of magnetization through the spin correlation ranks:

$$\begin{cases} \frac{\partial \rho(x,t)}{\partial t} = -2h\frac{\partial \rho(x,t)}{\partial x} - xr\rho(x,t) + r_0\delta(x-1) \\ \rho(x,0) = \delta(x-1), \quad \rho(\infty,t) = 0 \end{cases} \qquad (19)$$

where the modulus brackets have been dropped for clarity, $\delta(x)$ is the delta function and the discrete variable $k$ has now been replaced with a continuous variable $x$. Equation (19) is the well known dissipative transport equation. The source term and the initial condition reflect the fact that the simulation starts in the $\mathfrak{L}_1$ subspace, which is also replenished by relaxation.

There are two cases that we must analyze regarding Equation (19) – the case where the simulation is carried out for a long time and the state space restriction approximation is to remain valid *at all times,* and the case where the simulation is carried out for a finite time and the approximation is to stay valid *for the duration of the simulation*. The first case requires the $t \to \infty$ asymptotic solution to Equation (19), which is easily obtained:

$$\begin{cases} -2h\frac{\partial \rho(x)}{\partial x} - xr\rho(x) + r_0\delta(x-1) = 0 \\ \rho(\infty) = 0, \, \rho(x<0) = 0 \end{cases} \Rightarrow \quad \rho(x) = \left(\frac{r_0}{2h}e^{\frac{r}{4h}}H(x-1)\right)e^{-\frac{rx^2}{4h}} \qquad (20)$$

where $H(x)$ is the Heaviside step function. Note the fast decay of the norm (illustrated in Figure 3) as the spin correlation level gets higher. The requirement that the fraction of the magnetization leaking outside the restricted state space $x \leq k$ (where the cut-off level $k$ is specified by the user) not exceed the user-specified tolerance $\xi < 1$ is then fulfilled if

$$\int_k^\infty \rho(x) dx \Big/ \int_1^\infty \rho(x) dx < \xi \quad \Rightarrow \quad k > 2\sqrt{\frac{h}{r}} \mathrm{erfc}^{-1}\left(\xi \mathrm{erfc}\left(\frac{1}{2}\sqrt{\frac{r}{h}}\right)\right) \tag{21}$$

For a typical proton NMR simulation with an average *J*-coupling of 5 Hz and an average single-spin relaxation rate of 1 Hz this requires $k = 8$ for the asymptotic magnetization fraction in higher spin orders to be less than 1%. This estimate agrees perfectly with Figure 4 (which is discussed in detail in Section 3) and the empirical observations made when running the *Spinach* library[10]. In the case of the systems studied by Suter *et. al.*, where the relaxation rates scale as the square root of the coherence order[22-25], the asymptotic solution and the error bound take the form

$$\rho(x) = \left(\frac{r_0}{2h} e^{\frac{r}{3h}} H(x-1)\right) e^{-\frac{rx^{3/2}}{3h}} \qquad E_{1/3}\left(\frac{k^{3/2}r}{3h}\right) \Big/ E_{1/3}\left(\frac{r}{3h}\right) < \xi, \tag{22}$$

where the exponential integral function $E_n(x)$ is defined as:

$$E_n(x) = \int_1^\infty e^{-xt}/t^n \, dt. \tag{23}$$

The norm decay is still predicted to be super-exponential as a function of coherence rank. Even in the pessimistic case where the relaxation rate does not increase at all as a function of coherence order, we would still get:

$$\rho(x) = \left(\frac{r_0}{2h} e^{\frac{r}{4h}} H(x-1)\right) e^{-\frac{rx}{2h}} \qquad k > \frac{2h}{r} \ln(1/\xi) + 1 \tag{24}$$

in which the norm now decays exponentially as we move up the coherence ranks. Note that these estimates do not depend on the total number of spins in the system. Since the dimension of the restricted state space is polynomial in the total number of spins[1], this constitutes a formal proof that accurate spin dynamics simulations may indeed be performed in polynomial time.

Equations (20)-(24) define the applicability range for the state space restriction approximation. In all cases, the critical parameter is the ratio $h/r$ of the largest spin-spin coupling to the smallest relaxation rate. Systems in which the time scale of the relaxation processes is comparable to the time scale of the dynamics (*e.g.* liquid-state NMR and ESR systems) are therefore likely to be accurately described using low-order correlations. Systems with slow relaxation and fast dynamics (*e.g.* solid state NMR, particularly at low temperatures) would have a large $h/r$ ratio and consequently require larger basis sets.

Based on this reasoning the only class of problems in which high-order correlations cannot be ignored are those with an $h/r$ ratio large enough for the cut-off level $k$ to exceed the number of spins in the system. Several examples of medium-sized solid state NMR systems where the dynamics does clearly involve the entire state space have recently been published by Dumez and co-workers[6,7]. Some reduction techniques (conservation law analysis, zero track elimination, path tracing and destination state screening[2,8,26]) are still applicable to those systems, but polynomial scaling cannot be guaranteed.

The short-time approximation[12] was not necessary in the reasoning above, but we can obtain the short-time accuracy bound by noting that an extra cut-off would be imposed by the fact that the system would simply not have enough time to evolve into certain states to any significant extent. For this we require the general time-dependent solution to Equation (19), which is:

$$\rho(x,t) = \left( \frac{r_0}{2h} e^{\frac{r}{4h}} \left[ H(x-1) - H(x-2ht-1) \right] + e^{-\frac{r(x-2ht)^2}{4h}} \delta(x-2ht-1) \right) e^{-\frac{rx^2}{4h}} \quad (25)$$

for the case where the relaxation rates scale linearly with correlation order and

$$\rho(x,t) = \left( \frac{r_0}{2h} e^{\frac{r}{3h}} \left[ H(x-1) - H(x-2ht-1) \right] + e^{-\frac{r(x-2ht)^{3/2}}{3h}} \delta(x-2ht-1) \right) e^{-\frac{rx^{3/2}}{3h}}$$

$$\rho(x,t) = \left( \frac{r_0}{2h} e^{\frac{r}{2h}} \left[ H(x-1) - H(x-2ht-1) \right] + e^{-\frac{r(x-2ht)}{2h}} \delta(x-2ht-1) \right) e^{-\frac{rx}{2h}} \quad (26)$$

for the cases in which they scale as $\sqrt{k}$ and remain static respectively. In all three cases, the solution is zero outside the $x \in [1, 1+2ht]$ interval, meaning that the maximum upward "speed of travel" for the subspace norms is $2h$. Therefore, if the state space is restricted to $k$-spin orders, the result of the simulation up to $t = (k-1)/2h$ is expected to be accurate even in the absence of relaxation – this is also illustrated in Figure 3.

**3. Numerical examples**

From the perspective provided by the theoretical reasoning above, liquid-state NMR systems may be viewed as favourable – the time scale of relaxation processes (milliseconds to seconds) approximately matches the time scale of most experiments and the state space restriction approximation is therefore expected to work well. This is illustrated in Figure 4 for a pulse-acquire NMR experiment on the 22-spin system of strychnine: throughout the experiment, five-spin correlations are only marginally populated and higher correlations are absent; this agrees with the estimate given in the paragraph following Equation (21). Because BRW relaxation the-ory is accurate and well developed for liquid state NMR spin systems[15-17], the assumption made in Equation (16) about the linear growth of relaxation rates with coherence order may be tested by direct inspection – indeed, as Figure 2 (left panel) demonstrates, the growth is approximately linear up to six-spin correlations. The slight bend seen in the relaxation rate plot may be an indication that the asymptotic behaviour does have a square root dependence on the correlation order (as seen by Suter *et al.* in larger systems[22-25]). This cannot at the moment be tested computationally – the complexity of calculations including $N$ spins up to $k$-spin orders scales as $(4N)^k$; even Figure 4 takes several days on a modern supercomputer. Notably, a CSA-dominated relaxation superoperator (Figure 2, middle panel) does not exhibit this bend, indicating that square root scaling could be specific to dipolar relaxation processes.

A more sophisticated 2D NOESY NMR experiment on the same spin system (Figure 5) exhibits similar behaviour – to a reasonable accuracy, all spin correlations above five-spin orders can in this case be ignored as a consequence of their slow accumulation rate and fast relaxation. Figures 5 also illustrates the fact that the state space restriction approximation does in no way rely on the short-time evolution approximation – six-spin correlations do emerge in the system, but their amplitude is kept down by relaxation and therefore the state space restriction to five-spin orders would be accurate *at all times*, and not just for a brief initial period as the time-dependent perturbation theory would suggest.

As per Equations (21)-(24), spin systems with a larger $h/r$ ratio would require a larger basis set. This is illustrated with an ESR system in Figure 6 – both the couplings and the relaxation rates in the pyrene-dicyanobenzene radical pair are in the MHz range, but the dynamics is still confined with high accuracy to correlations of order below eight. Although the basis is larger than that of the previous example, it is still orders of magnitude smaller than the complete basis set. A match to the condition prescribed by Equation (21) is also very good in this case: with a typical (for organic radicals) hyperfine coupling of 10 Gauss and a typical relaxation rate of 10 MHz, Equation (21) requires spin orders up to 7 for the simulation to be accurate to 1%. Just as in the NMR example above, the assumption about the linear scaling of relaxation rates with the coherence order in liquid-state ESR systems is verified by direct inspection (Figure 2, right panel). While $O(N^7)$ scaling may appear steep, it is still fundamentally better than the exponential scaling of the brute-force calculation. A parallel may be drawn here to Quantum Chemistry and the relationship between CI-SDT (which is realistic), and full configuration interaction (which is not).

It is important to note that all cross-correlations have been kept in the relaxation superoperators used in the example calculations presented in Figures 4-6 – this implicitly validates our decision to ignore their presence during the derivation of the error bounds. When identical simulations are performed with cross-correlated relaxation terms zeroed out, there is no significant difference in the density operator norm behaviour from the one described above.

A solid state simulation example is given in Figure 7. Because solid state spin relaxation theories currently lack predictive power (as a consequence of the variable nature of spin-phonon coupling and energy spectrum of the phonon bath), the relaxation rates were set manually to the values reasonably expected for such systems from the Authors' practical experience (see the figure caption). As Figure 7 demonstrates, state space restriction appears to be a good approximation for solid effect DNP simulations. Further studies on restricted state space simulations of DNP are currently in progress and will be published by the Authors in due course.

Technical details of all numerical simulations (*Gaussian03* logs with the magnetic parameters and *Spinach* library console output) are available in the Supplementary Information[27].

## 4. Conclusions and outlook

The analysis of the complicated interplay of couplings and relaxation in the density matrix dynamics demonstrates that the state space restriction approximation in a suitably chosen basis set is likely to be applicable to many large spin systems, the primary reason being the inevitable presence of relaxation processes. The practical choice of basis for time domain simulations should be guided by Equations (21)-(24) and assumptions about the scaling law for the relaxation rates of high order spin coherences. We conclude that in many (and likely most) liquid state magnetic resonance systems state space restriction is a good approximation and a complete basis set is not required. In common with Quantum Chemistry, this does not necessarily mean that the simulation is going to be easy, but it does certainly mean that scaling would be polynomial, rather than exponential, in the number of spins.

**Acknowledgements**

The project is funded by the EPSRC (EP/F065205/1, EP/H003789/1, EP/I027254/1) and supported by the Oxford e-Research Centre.

**Figure captions**

**Figure 1** A schematic illustration of the flow of the density matrix norm according to Equation (13) in the subspace hierarchy given by Equation (2): $\mathfrak{L}_1$ corresponds to single-spin product states, $\mathfrak{L}_2$ to two-spin product states, *etc.* The individual subspaces $\mathfrak{L}_k$ are invariant under the action by $\hat{\hat{H}}_1$, but are coupled to nearest neighbours by $\hat{\hat{H}}_2$ and drained by the relaxation processes defined in $\hat{\hat{R}}_1$. The single-spin order subspace $\mathfrak{L}_1$ is also replenished by relaxation. Because the spin system dynamics starts in $\mathfrak{L}_1$ or $\mathfrak{L}_2$, very high levels in this subspace hierarchy could be left unpopulated if relaxation is fast enough to drain all the magnetization *en route* – upper bound estimates of the rates required are given in Equations (21)-(24).

**Figure 2** Norm of the projection of the Bloch-Redfield-Wangsness relaxation superoperator (including all cross-correlations) into the $k$-spin order subspace $\mathfrak{L}_k$ as a function of $k$ for: (*left*) a dipole coupling dominated 10-spin system of pyrene at 14.1 T with a rotational correlation time of 100 ps; (*centre*) a chemical shielding anisotropy dominated 10-spin system of perfluoropyrene at 14.1 T with a rotational correlation time of 100 ps; (*right*) a hyperfine coupling anisotropy dominated 11-spin system of perfluoropyrene cation radical at 0.33 T with a rotational correlation time of 10 ps. The relaxation superoperators were computed with the diagonalization-free relaxation theory module implemented in the *Spinach* library[10]. The geometries and interaction tensors were obtained from a GIAO DFT B3LYP/EPR-II calculation using the *Gaussian03* package (calculation logs available in the Supplementary Information[27]). It should be noted that the matrices in question are far too large to be diagonalized (in all three cases the dimension of $\hat{\hat{R}}_1$ exceeds 400,000) – the largest eigenvalue was calculated using power series analysis (as implemented in Matlab's *normest*) and the RMS eigenvalue was obtained from the Frobenius norm.

**Figure 3** An illustration of the upper bound accuracy estimate given in Equation (21). The presence of relaxation processes and finite amplitude of spin-spin couplings guarantee that a dynamic process starting in $\mathfrak{L}_1$ or $\mathfrak{L}_2$ would: (**A**) take time to reach higher correlation levels with the upward transport rate depending on the norm of

|            | $\hat{\hat{H}}_2$ and (**B**) populate higher correlation levels to a lesser extent, the fading rate being determined by the relative rates of upward transport and relaxation. |
|---|---|
| **Figure 4** | Numerical simulation of the density matrix norm dynamics during the evolution and detection period of a pulse-acquire NMR experiment on the 22-spin system of strychnine (technical details are available in the Supplementary Information[27]). All distances and magnetic parameters were imported from a GIAO DFT B3LYP/EPR-II calculation. A Bloch-Redfield-Wangsness relaxation superoperator (including DD, CSA and cross-correlation terms) was used with an isotropic rotational diffusion correlation time of 200 ps. |
| **Figure 5** | Numerical simulation of the density matrix norm dynamics during the detection period of the NOESY experiment on the 22-spin system of strychnine (technical details are available in the Supplementary Information[27]). All distances and magnetic parameters were imported from a GIAO DFT B3LYP/EPR-II calculation. A Bloch-Redfield-Wangsness relaxation superoperator (including DD, CSA and cross-correlation terms) was used with an isotropic rotational diffusion correlation time of 200 ps. |
| **Figure 6** | Numerical simulation of the density matrix norm dynamics during the evolution of a singlet-born pyrene-dicyanobenzene radical pair at a magnetic induction of 10 Gauss (technical details are available in the Supplementary Information[27]). The Hore-Jones radical recombination kinetics superoperator was used with a singlet recombination rate of $4 \cdot 10^7$ s$^{-1}$. All magnetic parameters were imported from a GIAO DFT B3LYP/EPR-II calculation; Bloch-Redfield-Wangsness relaxation theory (with $\hat{H}_0$ containing the isotropic Zeeman and hyperfine interactions) was used to obtain the relaxation superoperator. |
| **Figure 7** | Numerical simulation of the density matrix norm dynamics during a solid effect DNP simulation of a system with a single electron and 7 nuclei located randomly at distances between 12Å and 18Å from the electron (technical details are available in the Supplementary Information[27]). Magnetic induction is 3.4 Tesla. Microwave irradiation is applied with a strength of 1.5 MHz at the electron-nuclear zero-quantum transition frequency. The longitudinal and transverse relaxation |

rates of the nuclei are 0.01 Hz and 10 Hz respectively. The longitudinal and transverse relaxation rates of the electron are $10^3$ Hz and $10^6$ Hz respectively.

figure 1

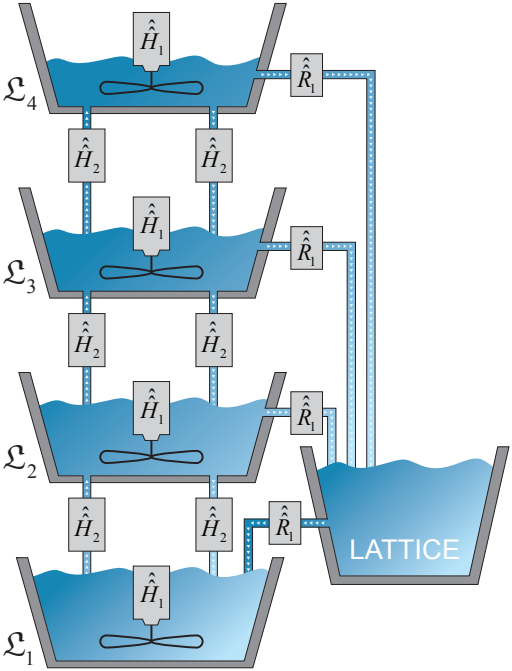

figure 2

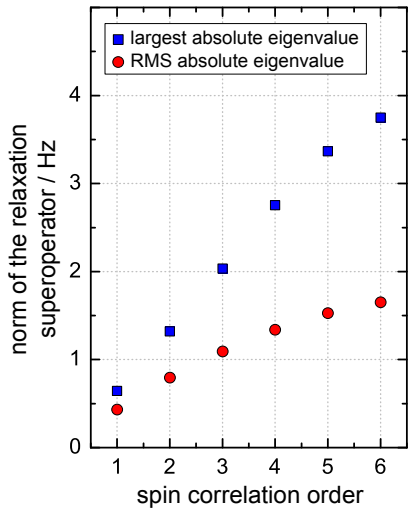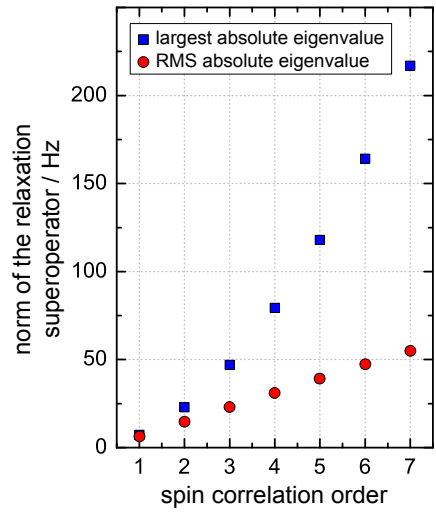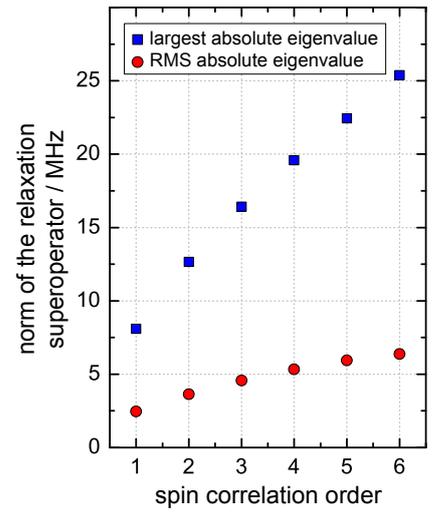

figure 3

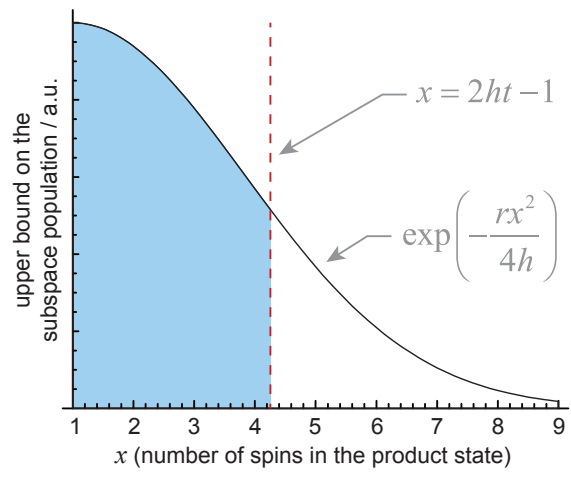

figure 4

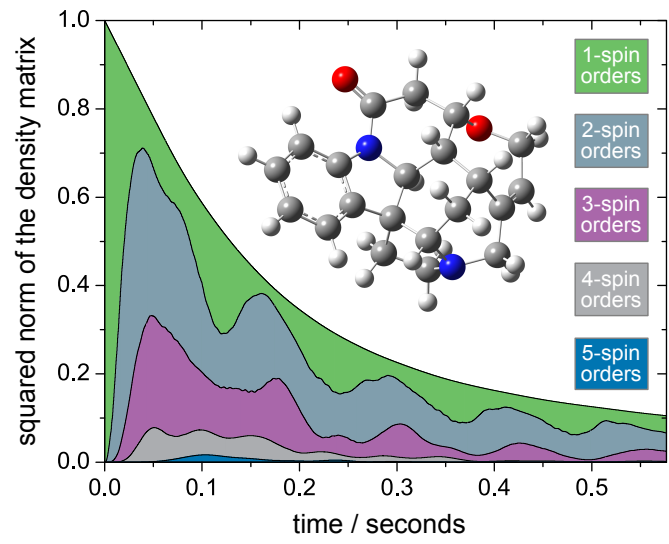

figure 5

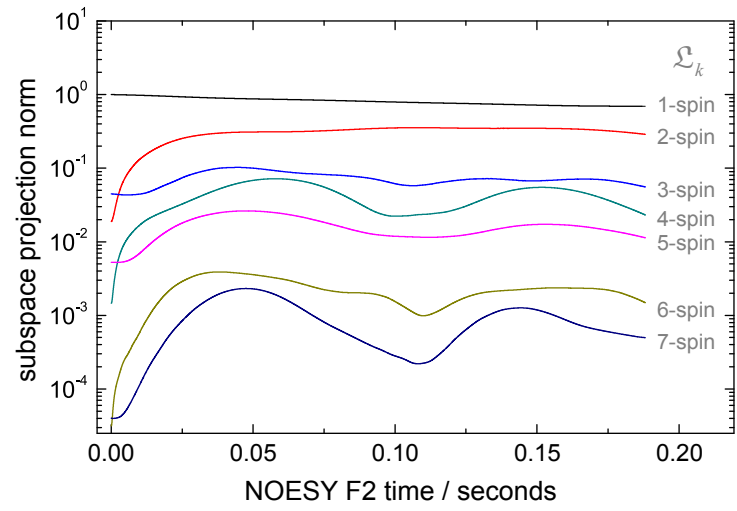

figure 7

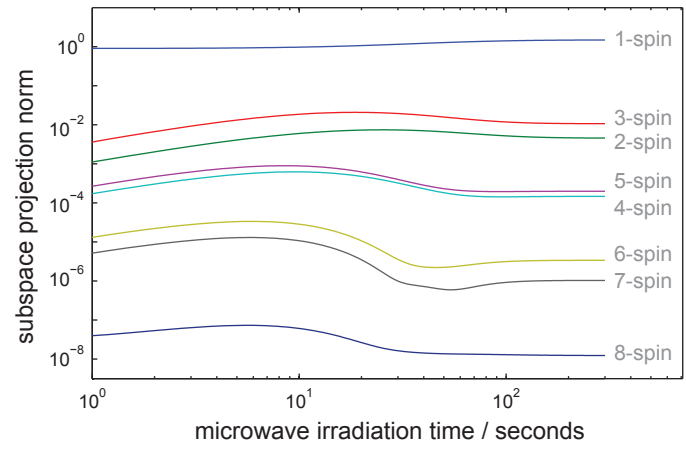